\documentclass{iau}
\usepackage{graphicx}

\usepackage{natbib}
\newcommand{\Oii}{\ensuremath{\mathrm{[O\,\textsc{ii}]}\ \lambda 3726,3729}}
\newcommand{\Neiii}{\ensuremath{\mathrm{[Ne\,\textsc{iii}]}\ \lambda 3869}}

\newcommand{\OII}{\ensuremath{\mathrm{[O\,\textsc{ii}]}}}
\newcommand{\NeIII}{\ensuremath{\mathrm{[Ne\,\textsc{iii}]}}}

\title[Nature and physical properties of ALMA selected galaxies using MUSE
spectroscopy] %
{Nature and physical properties of gas-mass selected galaxies using integral
  field spectroscopy}

\author[Leindert A. Boogaard]   %
{Leindert A. Boogaard$^1$}

\affiliation{$^1$Leiden Observatory,
  Leiden University, \\ PO Box 9513, NL-2300 RA Leiden, The Netherlands \\ email: {\tt boogaard@strw.leidenuniv.nl}} %

\pubyear{2019}
\volume{352}  %
\jname{Uncovering early galaxy evolution in the ALMA and JWST era}
\editors{}%
\begin{document}

\maketitle

\begin{abstract}
  Mapping the molecular gas content of the universe is key to our understanding
  of the build-up of galaxies over cosmic time.  Spectral line scans in deep
  fields, such as the Hubble Ultra Deep Field (HUDF), provide a unique view on
  the cold gas content out to high redshift.  By conducting
  `spectroscopy-of-everything', these flux-limited observations are sensitive
  to the molecular gas in galaxies without preselection, revealing the cold gas
  content of galaxies that would not be selected in traditional studies.

  In order to capitalize on the molecular gas observations, knowledge about the
  physical conditions of the galaxies detected in molecular gas, such as their
  interstellar medium conditions, is key.  Fortunately, deep surveys with
  integral-field spectrographs are providing an unprecedented view of the
  galaxy population, providing redshifts and measurements of restframe
  UV/optical lines for thousands of galaxies.

  We present the results from the synergy between the ALMA Spectroscopic Survey
  of the HUDF (ASPECS), with deep integral field spectroscopy from the MUSE
  HUDF survey and multi-wavelength data.  We discuss the nature of the galaxies
  detected in molecular gas without preselection and their physical properties,
  such as star formation rate and metallicity.  We show how the combination of
  ALMA and MUSE integral field spectroscopy can constrain the physical
  properties in galaxies located around the main sequence during the peak of
  galaxy formation.

\end{abstract}

\firstsection %
\section{Introduction}
Recent years have seen tremendous advances in the characterization of the
cosmic history of star formation and it has now been established that the star
formation rate (SFR) density increased with cosmic time up to a peak at
$z\sim1-3$ and then decreased until the present \citep[for an overview,
see][]{Madau2014}.  At each epoch, more massive star-forming galaxies are
observed to have a higher star formation rate, establishing what has become
known as the `galaxy main sequence' \citep[MS;][]{Brinchmann2004, Noeske2007a,
  Whitaker2014, Schreiber2015, Boogaard2018}.

A key ingredient in our understanding of galaxy formation is the observation of
the cold interstellar medium (ISM) -- the `fuel for star formation' --
typically traced through carbon monoxide ($^{12}$CO, hereafter CO) or dust
continuum emission \citep[for a review, see][]{Carilli2013}.  Recent years have
seen a significant progress in observations of CO at $z>1$, through targeted
observations of star-forming galaxies on and above the MS
\citep[e.g.,][]{Daddi2015, Tacconi2018, Silverman2018}.  This has been largely
driven by the incredible sensitivity of the Atacama Large Millimeter Array
(ALMA), which is now revolutionizing the field of high-redshift ISM line
observations, as shown by several publications in this volume.

In order to conduct a `complete' census of the cosmic molecular gas density,
one has to probe a well defined cosmic volume without any target preselection.
So called `spectral scan surveys' are designed to do exactly that, by
conducting a mosaic of observations on the sky while simultaneously scanning a
large bandwidth in frequency for emission lines from the cold ISM.  These are
now providing the first ever constraints on the cosmic molecular gas content
and hence the cosmic star formation efficiency, directly from CO observations
\citep{Decarli2014, Walter2014, Walter2016, Pavesi2018, Riechers2019,
  Decarli2019}.

As molecular line emission can be elusive, having a large number of
spectroscopic redshifts over the target field can provide essential redshift
information to push spectral scans to their sensitivity limit.  The Multi-Unit
Spectroscopic Explorer (MUSE) on the Very Large Telescope \citep{Bacon2010} is
a key instrument in this context, providing `integral field spectroscopy' over
a large $1' \times 1'$ field-of-view from $4750 - 9300$~\AA\ at a high
observing efficiency.

Here, we will present the first results from the ALMA Spectroscopic Survey
(ASPECS) in the Hubble Ultra Deep Field (HUDF) Large Program
\citep{Decarli2019}, focusing specifically on the synergy between ALMA and MUSE
\citep{Boogaard2019}.  These results are part of a larger series of papers on
the Band 3 data from ASPECS, further discussed in \cite{Gonzalez-Lopez2019,
  Popping2019, Aravena2019}.

\begin{figure}[t]
  \begin{center}
    \includegraphics[width=3.4in]{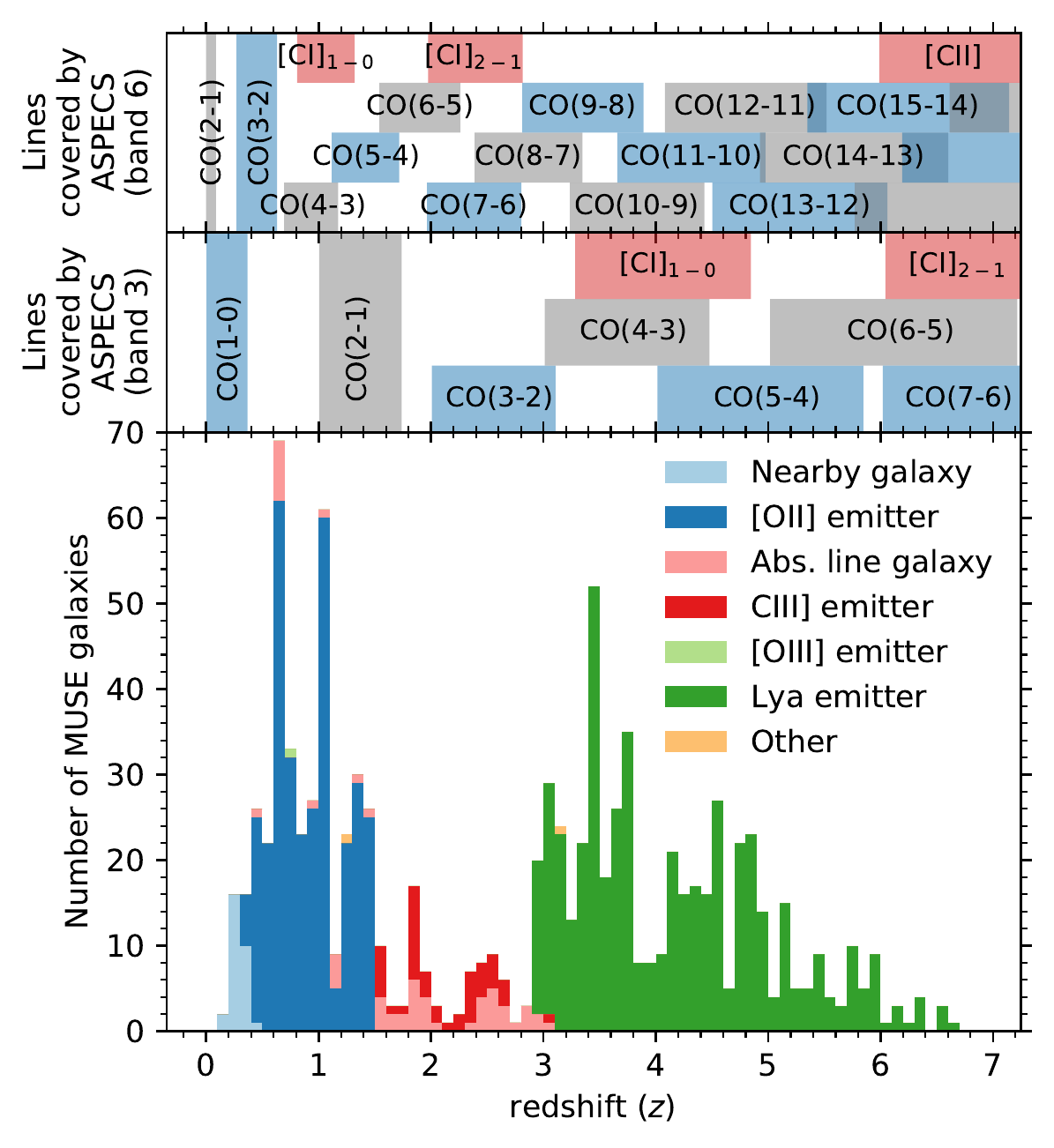}
    \caption{\textbf{Top panel:} Primary emission lines observable at different
      redshifts with the ALMA Spectroscopic Survey (ASPECS) in Band 3 and 6.
      \textbf{Bottom panel}: Histogram of all galaxies from the MUSE HUDF
      Survey \citep{Bacon2017, Inami2017} that fall within the ASPECS field
      \citep{Decarli2019}, colored by their spectral classification.  At each
      distinct redshift, MUSE and ALMA probe different parts of the galaxy
      spectra. \emph{Figure adapted from \cite{Boogaard2019}.}}
    \label{fig:muse-aspecs-histogram}
  \end{center}
\end{figure}

\section{Observations: ALMA and MUSE observations in the HUDF}
The ALMA Spectroscopic Survey consists of a mosaic of spectral scan
observations targeting the deepest region of the HUDF, covering the complete
ALMA Band 3 (84 - 115 GHz) and Band 6 (212 - 272 GHz) windows.  These
flux-limited observations allow one to detect both line emission from the ISM
(mainly CO, [CI] and [CII]) as well as dust continuum at different redshifts,
without any preselection of targets (see the top panels of
Fig.~\ref{fig:muse-aspecs-histogram}).  The HUDF is an excellent target for
ASPECS, due to the wealth of multi-wavelength data available all the from the
X-ray to radio.  In particular, deep \emph{HST} and \emph{Spitzer} photometry
are key to identify the host galaxies of the molecular line emission, as the
stellar light from these sources is often significantly attenuated.

The MUSE data over the HUDF provides optical spectroscopy for all galaxies in
the field and has revolutionized the amount of available redshift information
(increasing the number of known spectroscopic redshifts by a factor
$\times 10$; \citealt{Bacon2017, Inami2017}).
Fig.~\ref{fig:muse-aspecs-histogram} shows how the galaxies at different
redshifts are identified by distinct spectral features that fall within the
spectrograph.

Fig.~\ref{fig:muse-aspecs-histogram} also illustrates how the integral field
data from ALMA and MUSE can work together: On the one hand, MUSE can provide
redshift information for the galaxies detected in CO emission as well as
physical properties from the rest-frame UV/optical spectra (depending on the
redshift).  On the other hand, one can leverage the large number of MUSE
redshifts to search specifically for lines in the ALMA data, as well as conduct
stacking in order to detect fainter sources.

\begin{figure}[t]
  \begin{center}
    \includegraphics[width=0.49\textwidth]{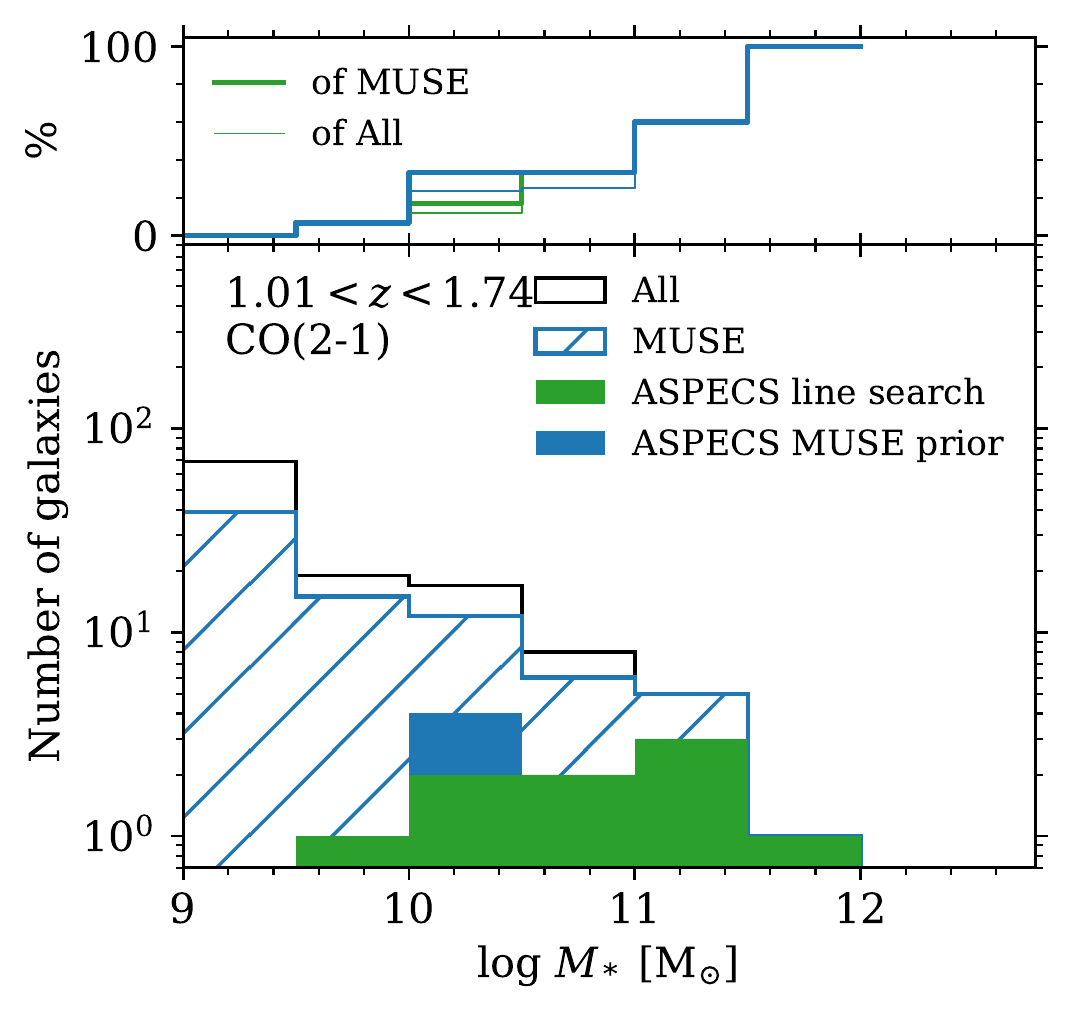}
    \includegraphics[width=0.49\textwidth]{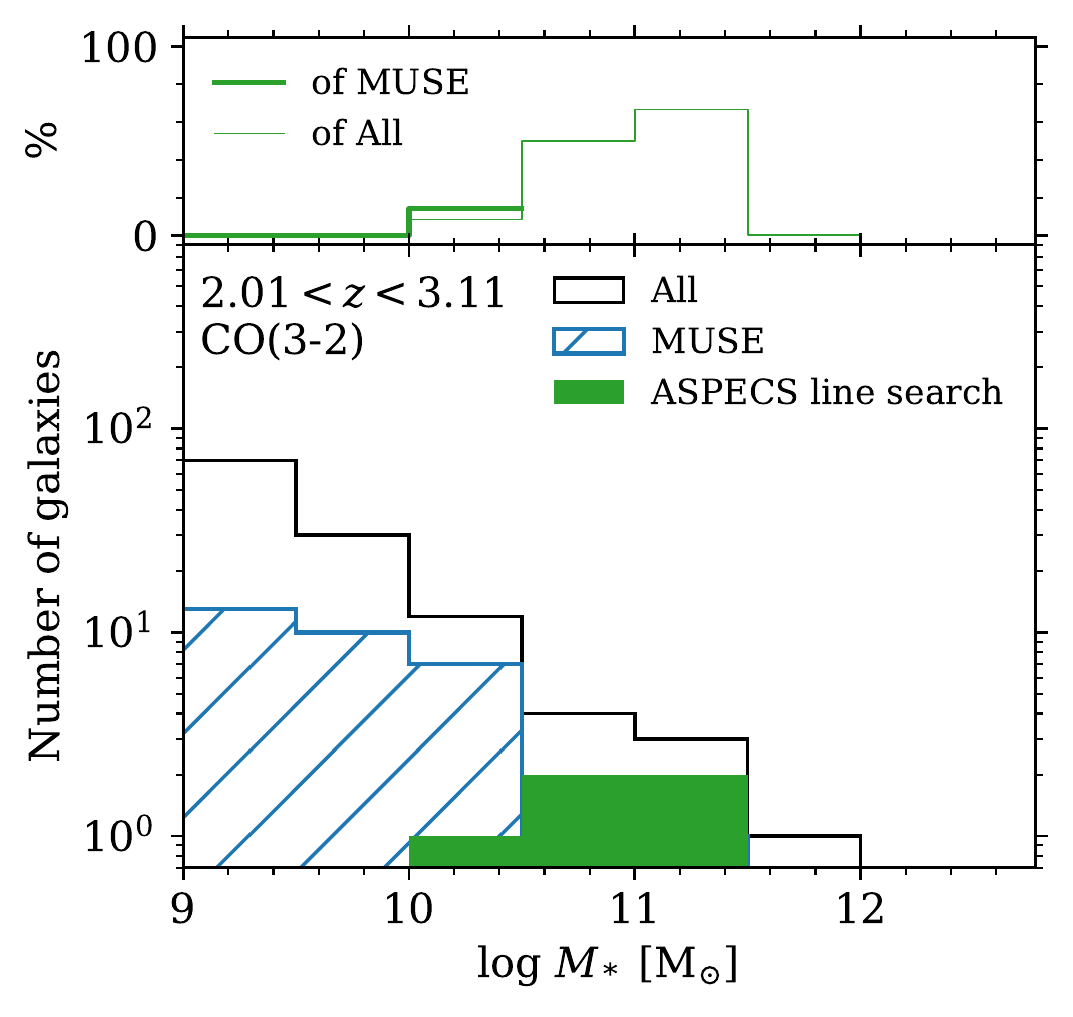}
    \caption{Detection fractions and histograms as a function of the stellar
      masses ($M_{*}$) of the ASPECS galaxies detected in CO(2-1) (\textbf{left
        panel}) and CO(3-2) (\textbf{right panel}).  With a purely gas-mass
      selected sample we detect most of the massive galaxies, while probing
      down to $M_{*} \approx 10^{9.5}$~M$_{\odot}$. \emph{Figure adapted from
        \cite{Boogaard2019}}}
    \label{fig:hist-mass}
  \end{center}
\end{figure}

\section{Results: The physical properties of the ASPECS galaxies}
\begin{figure}[t]
  \begin{center}
    \includegraphics[width=0.46\textwidth]{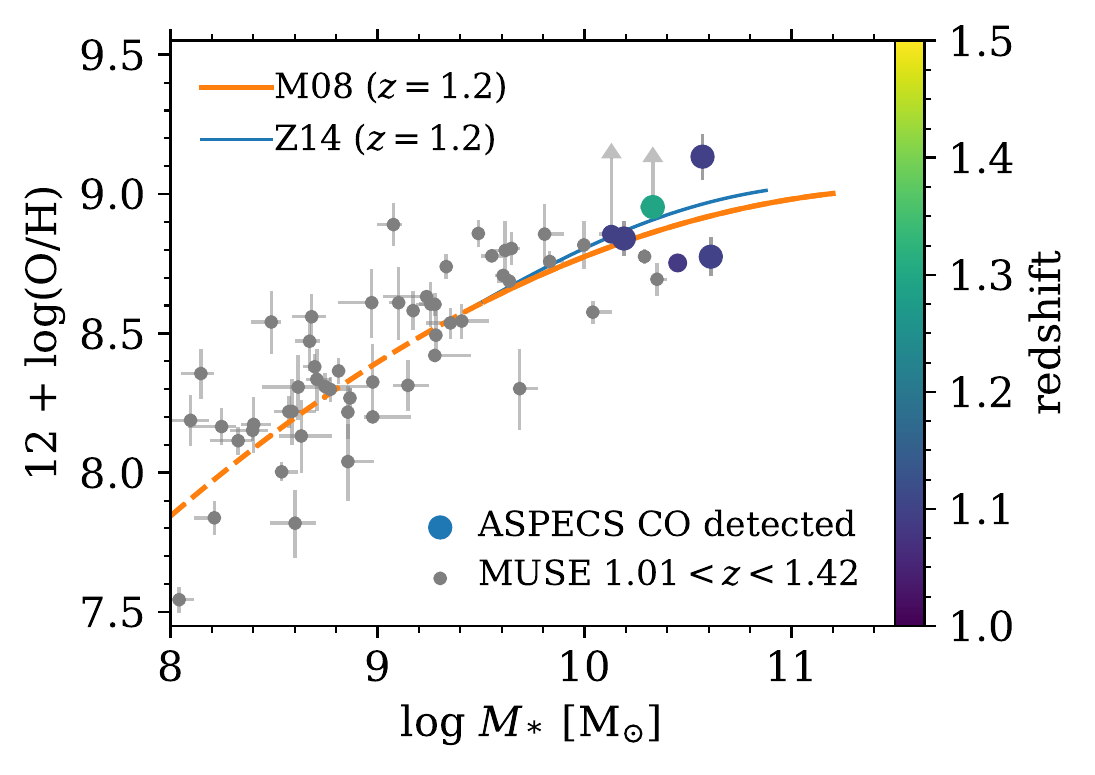}
    \includegraphics[width=0.49\textwidth]{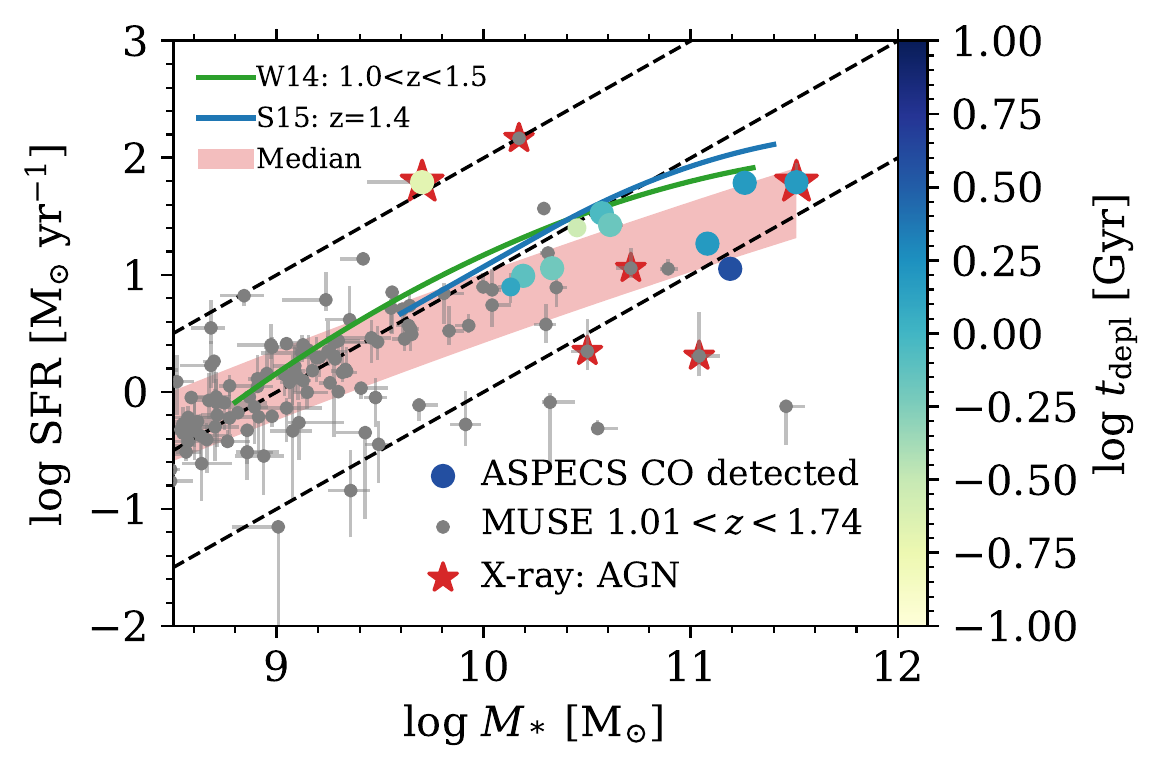}
    \caption{\textbf{Left panel:} Stellar mass - metallicity relation of the
      ASPECS galaxies detected in CO(2-1), derived from the \OII/\NeIII\ ratio
      (at $z<1.42$).  The ASPECS galaxies are consistent with a (super-)solar
      metallicity.  \textbf{Right panel:} Stellar mass - star formation rate
      relation for the ASPECS galaxies detected in CO(2-1).  With ASPECS with
      detect galaxies on, above and below the `galaxy main sequence' at these
      redshifts. \emph{Figures adapted from \cite{Boogaard2019}}}
    \label{fig:msfr-mmol}
  \end{center}
\end{figure}
We search the complete ASPECS Band 3 cube for emission lines using a matched
filtering approach \citep{Gonzalez-Lopez2019}.  This reveals 16 emission line
candidates at high significance (very low probability of being spurious), all
of which show a counterpart in the \emph{HST} imaging.  Using the MUSE and
multi-wavelength data, we identify all as rotational transitions of CO, with
redshifts between $z=1.0 - 3.6$ ($J_{\rm up} = 2 - 4$).  In addition, using the
MUSE redshifts as prior information, we recover two additional CO lines at
signal-to-noise $>3$, bringing the total sample to 18.  We convert the observed
CO luminosity to CO(1-0) using the excitation corrections derived for
$z\sim1.5$ star-forming galaxies from \cite{Daddi2015} and convert this to a
molecular gas mass assuming an $\alpha_{\rm CO} = 3.6$ \citep{Daddi2010},
similar to the Galactic value \citep[e.g.,][]{Bolatto2013}.

For the galaxies detected in CO(2-1) up to $z=1.5$, MUSE still covers part of
the NUV spectrum (see Fig.~\ref{fig:muse-aspecs-histogram}), where we can use
emission line ratios to probe the metallicity.  We use the ratio of
\Oii/\Neiii\ to infer the gas-phase metallicity in these galaxies, following
the calibration from \cite{Maiolino2008}, as shown in the left panel of
Fig.~\ref{fig:msfr-mmol}.  Overall, we find that the ASPECS galaxies have a
(super-)solar metallicity, consistent with what would be expected from the
mass-metallicity relation \citep[e.g.][]{Zahid2014}, supporting the use of a
Galactic $\alpha_{\rm CO}$ for these galaxies.

A key question is what kind of galaxies we detect with the flux-limited
observations from ASPECS.  In Fig.~\ref{fig:hist-mass}, we plot histograms of
the stellar masses of the galaxies in which we detect molecular gas.  What is
immediately evident is that we detect molecular gas in the majority of the most
massive galaxies (in terms of their stellar mass).  As we move towards lower
mass galaxies, the fraction of galaxies decreases, reaching a detection
fraction of $\sim$50\% for galaxies with $M_{*} \ge 10^{10}$ ($10^{10.5}$)
M$_{\odot}$ at $1 < z < 2$ ($2 < z < 3$).

In the right panel of Fig.~\ref{fig:msfr-mmol}, we expand the one dimensional
histograms and show the ASPECS galaxies on the stellar mass - SFR plane.  At
$1<z<2$, we find that our gas-mass selected sample consists mostly of galaxies
on the `galaxy main sequence' at these redshifts.  Above
$M_{*} \approx 10^{10}$~M$_{\odot}$, we detect almost all of the `starburst'
galaxies, defined as lying $\ge 0.3$~dex above the main sequence.  Critically,
we also detect galaxies that lie below the MS.  While these galaxies have SFR
that is lower than typical at their stellar mass, they still host a significant
gas reservoir.  Their detection in the flux-limited survey is important, as
these galaxies would typically not be selected in targeted follow-up.

The detections fractions as a function of mass and SFR, together with
additional CO lines found using the MUSE redshifts as a prior, suggest that
there may be several galaxies for which the CO emission falls just below the
individual detection threshold.  The large number of systemic redshifts from
MUSE can be leveraged here, in order to recover this signal through stacking.
Indeed, preliminary results show that stacking the CO(2-1) undetected galaxies
on the MS at $10 < \log M_{*} [\mathrm{M}_{\odot}] < 11$ show a detection of
their CO line emission.  These and additional results from the stacking will be
discussed in Inami et al. (in prep.).

\section{Outlook: ASPECS and \emph{JWST}}
The combination of integral field spectroscopy with MUSE in the optical and
ALMA in the (sub-)millimeter regime provides a unique tool for studying the
physical properties of galaxies across cosmic time.  The recently observed band
6 data from ASPECS provide additional constraints on the molecular gas content,
through the deep dust continuum map at 1.2~mm and observations of atomic
carbon, as well as on the CO excitation, by constraining higher-$J$ CO lines
for all sources.

The anticipated launch of the \emph{James Webb Space Telescope} will open up a
new window on these galaxies.  In particular, key rest-frame UV, optical and
near-infrared spectroscopy will allow further characterization of the star
formation rates, metallicities and ISM conditions in these galaxies.  Together
with observations of the cold ISM from ALMA, these will provide critical
constraints on our theory of star formation at high redshift.

\section*{Acknowledgements}
The work presented here is the result of a large collaborative effort and would
not have been possible without Fabian Walter, Roberto Decarli, Manuel Aravena,
Jorge Gonz\'alez-L\'opez, Chris Carilli, Paul van der Werf, Rychard Bouwens,
Roland Bacon, Hanae Inami, and the other members from the ASPECS and MUSE GTO
teams.  The author would like to thank Paul van der Werf for providing comments
on the manuscript.

\end{document}